\pdfoutput=1
\documentclass{llncs}
\usepackage[english]{babel}
\usepackage[utf8]{inputenc}
\usepackage{amsmath}
\usepackage{amsfonts}
\usepackage{amssymb}
\usepackage{graphicx}
\usepackage{hyperref}

\usepackage{times}
\usepackage{fancybox}
\usepackage{xspace}
\usepackage{cleveref}
\usepackage{cite}
\usepackage{caption}
\usepackage[cache=true]{minted}
\usepackage[position=below]{subcaption}
\usepackage{tikz}
\usetikzlibrary{calc,shapes,chains,decorations.markings,positioning,arrows,fit,decorations,backgrounds}  
\usepackage{pgfplots}
\usepackage[text size=small, backgroundcolor=blue!30,bordercolor=blue,linecolor=blue]{todonotes}

\DeclareCaptionSubType[alph]{listing}

\crefname{listing}{Listing}{Listings}
\crefname{sublisting}{Listing}{Listings}
\crefname{figure}{Figure}{Figures}
\crefname{subfigure}{Figure}{Figures}

\def\figurename{Figure}

\newmintinline[eiffel]{eiffel}{fontsize=\small}
\newmintinline[java]{java}{fontsize=\small}

\newminted[eiffelcode]{eiffel}{fontsize=\scriptsize,tabsize=4}
\newminted[javacode]{java}{fontsize=\scriptsize,tabsize=4}

\newcommand{\scoop}{\textsc{Scoop}\xspace}
\newcommand{\dscoop}{\textsc{D-Scoop}\xspace}
\newcommand{\msg}[1]{\ovalbox{\scriptsize{#1}}}

\newcommand{\myparagraph}[1]{\smallskip\noindent\textbf{#1}\hspace{1.5ex}}

\title{An Interference-Free Programming Model\\ for Network Objects}
\author{Mischael Schill\inst{1} \and Christopher M. Poskitt\inst{2} \and Bertrand Meyer\inst{3,4,5}}
\institute{Department of Computer Science, ETH Z\"{u}rich, Switzerland \and%
  Singapore University of Technology and Design, Singapore \and%
  Politecnico di Milano, Italy\and%
  Innopolis University, Russia\and%
  Universit\'{e} Paul Sabatier, France}

\begin{document}
\maketitle
\begin{abstract}
	Network objects are a simple and natural abstraction for distributed object-oriented programming. Languages that support network objects, however, often leave synchronization to the user, along with its associated pitfalls, such as data races and the possibility of failure. In this paper, we present \mbox{\dscoop}, a distributed programming model that allows for interference-free and transaction-like reasoning on (potentially multiple) network objects, with synchronization handled automatically, and network failures managed by a compensation mechanism. We achieve this by leveraging the runtime semantics of a multi-threaded object-oriented concurrency model, directly generalizing it with a message-based protocol for efficiently coordinating remote objects. We present our pathway to fusing these contrasting but complementary ideas, and evaluate the performance overhead of the automatic synchronization in \mbox{\dscoop}, finding that it comes close to---or outperforms---explicit locking-based synchronization in Java RMI.
\end{abstract}
\section{Introduction}\label{sec:introduction}

Inter-device communication is becoming ubiquitous, and the number of connected devices is growing everyday. With this ubiquity comes an increasing demand for programmers to be able to write reliable distributed software, yet this is no simple task. Challenging errors such as data races and deadlocks can arise from subtle mistakes in synchronization code; and the failure of individual devices can block whole systems in the absence of appropriate recovery protocols.

Various language abstractions have been proposed to make it easier to write distributed programs. One such abstraction, natural for the object-oriented paradigm, is that of \emph{network objects}~\cite{Birrell:1993}: objects whose methods can be invoked over a network. By handling communication in method calls, network objects allow for local and remote objects to be treated uniformly, without regard to where they are physically located. In principle an elegant generalization; in practice, languages supporting them are often lightweight on synchronization, leaving the user to manage it explicitly, and potentially exposing them to the aforementioned errors.

Many of these pitfalls of synchronization are not unique to distribution: they occur in multi-threaded concurrent programming too. Several languages and libraries attempt to make it easier and safer to write concurrent programs, providing their users with high-level abstractions as diverse as transactional memory~\cite{Shavit-Touitou97a}, block-dispatching~\cite{GCD-Reference}, actors~\cite{Agha86a}, and active objects~\cite{Lavender-Schmidt96a}. Given the many shared synchronization challenges, a number of these abstractions have been successfully applied across novel distributed programming approaches, exemplified by languages such as Creol~\cite{Johnsen-Owe-Yu06a}, JCoBox~\cite{Schaefer-PH10a}, and AmbientTalk~\cite{Dedecker-et_al06a}.

A family of concurrency abstractions that (until the present paper) had not been generalized to distributed programming were those provided by \scoop~\cite{West-NM15b}, despite their potential to naturally complement the network objects abstraction and to address some of its shortcomings. \scoop is an object-oriented concurrency model that provides data-race freedom by construction, and strong guarantees about the order in which requests are executed by concurrently running processes. The synchronization provided by its runtime automatically excludes interfering calls, making it possible to reason independently about different blocks of code over multiple concurrent objects, almost as if each block is ``sequential''. The ethos of the \scoop approach---stick to the mental models programmers already know well (in this case sequential programming)---is aligned with that of the network objects abstraction, and challenged us to explore how they could complement the strengths of each other.

\myparagraph{Our Contributions.} The main outcome of this paper is \dscoop, a distributed programming model resulting from the fusion of the network objects abstraction with the runtime of the \scoop concurrency model. The strong reasoning guarantees of the latter are directly generalized to provide interference-free and transaction-like reasoning on (potentially multiple) network objects, without the programmer having to worry about how to achieve it. The basis of this fusion is a message-based protocol for coordinating remote objects, which includes an efficient and novel two-phase locking algorithm for establishing the \scoop order guarantees without prolonged periods of blocking. Furthermore, we adapt from transactional memory the recovery technique of compensations, in order for \dscoop to be able to restore consistency when clients fail mid-computation. This paper presents our pathway to fusing these independent, but complementary ideas. We furthermore evaluate a prototype implementation of 
\dscoop to investigate the performance overhead of its automatic synchronization mechanisms, finding that they come close to---and in some circumstances outperform---explicit locking-based synchronization in the Java RMI realization of network objects.

For the distributed programming community, this paper presents a programming model with interference-free and transaction-like reasoning for distributed objects, and a runtime that effectively handles the synchronization. For the \scoop community, it presents a generalization of the classical \scoop concurrency model to distribution in a way that maintains the guarantees of the core abstractions. For language designers, it presents a simple yet effective distributed programming abstraction (and descriptions of how we realized it) that could be transferred to other object-oriented languages.

\myparagraph{Plan of the Paper.}
After introducing the necessary technical background of network objects and \scoop (\Cref{sec:scoop}), we show how they fuse together in \dscoop, our distributed programming model (\Cref{sec:architecture}). We go into more depth on how objects are controlled to avoid interference (\Cref{sec:control}) and how compensation helps in managing failure (\Cref{sec:compensation}). Our prototype is then evaluated against Java RMI (\Cref{sec:evaluation}), before we review some related work (\Cref{sec:related}) and conclude (\Cref{sec:conclusion}).

\section{Background: Network Objects and \scoop}\label{sec:scoop}

Our work combines networks objects---a distributed programming abstraction---with \scoop, a concurrency model that handles synchronization in its runtime and provides strong reasoning guarantees. We present the necessary technical background of these concepts in the context of a running example.

\myparagraph{Network Objects.} A \emph{network object} is an object whose methods can be invoked over a network. The abstraction is a simple but natural generalization of standard objects to distributed contexts: the programmer interacts with their interfaces in the same way as before, and without regard to where the object is physically located. Communication is handled in the method calls, and is typically synchronous to mimic regular method calls. Network objects first appeared in Modula-3~\cite{Birrell:1993}, and have since strongly influenced Java's Remote Method Invocation (RMI) API as well as the Common Object Request Broker Architecture (CORBA) standard.

While implementations of network objects vary, the abstraction is typically light\-weight on synchronization, leaving this difficulty to the user, to the point that multiple clients can concurrently execute the same method (introducing the possibility of data races). Simple mechanics such as \java{synchronized} in Java are not always sufficient to ensure atomicity. Consider for example the simple bank account \eiffel{transfer} method in \cref{fig:transfer}, which allows some client to transfer an amount (\eiffel{am}) of money from a source (\eiffel{s}) account to a target (\eiffel{t}) account. If the system is single-threaded and the accounts are local, then the method is correct. If the accounts can be accessed concurrently, then locks or other measures are required to ensure the atomicity of \eiffel{transfer}. If however the accounts are remote and can be accessed concurrently as network objects, then we must adapt again.

\begin{listing}[htb]
	\begin{subfigure}[t]{0.5\textwidth}
		\begin{eiffelcode}
transfer (s, t: ACCOUNT; 
		am: NATURAL)
do
	if s.balance >= am then
		s.set_balance (s.balance - am)
		t.set_balance (t.balance + am)
	else
		-- Notify user
	end
end
		\end{eiffelcode}
	\end{subfigure}
	\begin{subfigure}[t]{0.5\textwidth}
		\begin{eiffelcode}
transfer (s, t: separate ACCOUNT;
		am: NATURAL)
do
	if s.balance >= am then
		s.set_balance (s.balance - am)
		t.set_balance (t.balance + am)
	else
		-- Notify user
	end
end
		\end{eiffelcode}
	\end{subfigure}
	\caption{Bank account transfer methods: sequential (left) and in \scoop (right)}
	\label{fig:transfer}
\end{listing}

One solution is to use locks and expose them as network objects, but this poses risk, e.g.~if a client loses its connection before having a chance to release its locks. Another solution is to hide the synchronization within additional methods in the account class, but this is still challenging to implement without introducing concurrency errors such as races or deadlocks. Either way, the simplicity of the network object abstraction suffers with the complexity of synchronizing correctly; hence our aim to elegantly integrate it with a concurrency model that can manage such complexity in its runtime. 

\myparagraph{\scoop.} \scoop \cite{West-NM15b} is a concurrent object-oriented programming model that aims to preserve the well-understood modes of reasoning enjoyed by sequential programs, such as pre- and postcondition reasoning over blocks of code. Programmers are provided with simple abstractions for expressing concurrency, with the runtime itself responsible for correctly handling synchronization. We describe \scoop in the context of its principal implementation for Eiffel~\cite{SCOOP-EiffelStudio-Reference}, but remark that the ideas generalize to other object-oriented languages (e.g.~Java~\cite{Torshizi-OPC09a}).

In \scoop, every object is associated with a \emph{process} (which we call its \emph{handler}), a concurrent thread of execution with the exclusive right to call methods on the objects it handles. In this context, object references may point to objects with the same handler (\emph{non-separate} objects) or to objects with distinct handlers (\emph{separate} objects). Method calls on non-separate objects are executed immediately by the shared process. To make a call on a separate object, however, a \emph{request} must be sent to the handler of that object to process it: if the method is a \emph{command} (i.e.~it does not return a result) then it is executed asynchronously, leading to concurrency; if it is a \emph{query} (i.e.~a result is returned and must be waited for) then it is executed synchronously. Note that processes cannot synchronize via shared memory: only by exchanging requests.

The possibility for objects to have different handlers is captured in the type system by the keyword \eiffel{separate}. To request method calls on objects of \eiffel{separate} type, programmers simply make the calls within \emph{separate blocks}: these are the bodies of any methods that have separate objects as formal parameters. \scoop provides guarantees about the order in which calls in these blocks are executed, so as to help programmers avoid concurrency errors. In particular, method calls on separate objects will be logged as requests by their handlers in the order that they are given in the program text; furthermore, there will be no intervening requests logged from other handlers. These guarantees exclude data races by construction, and allow programmers to apply sequential reasoning within separate blocks independently of the rest of the program.

Consider the concurrent version of \eiffel{transfer} in \cref{fig:transfer}, in which bank account objects have concurrently running handlers. Suppose that a process calls the method \eiffel{transfer (acc1, acc2, 100)} on \eiffel{separate} accounts \eiffel{acc1} and \eiffel{acc2}. The body of the method contains two commands on these \eiffel{separate} objects---thus, two asynchronously executed requests---that transfer the stated amount from the first account to the second. It also contains \eiffel{balance} queries which are executed synchronously. The \scoop guarantees ensure that while the process is inside the body of \eiffel{transfer}, no other process can log intervening requests on \eiffel{acc1} or \eiffel{acc2}. As a result, it would not be possible for another process to observe the balances of the two accounts in an intermediate state, i.e.~when the money has been withdrawn from the former but not credited to the latter. The body of \eiffel{transfer} can thus be reasoned about sequentially and independently of the rest of the program. This additional control over the order in which requests are logged (i.e.~that requests cannot be interrupted) is the key distinction \scoop has over other message-passing-based models such as the actor model, or active objects.

\scoop provides some more advanced concurrency mechanisms beyond the focus of this paper. Most notable are its generalization of method preconditions to support condition synchronization on \eiffel{separate} objects, and its support for efficient data sharing between processes sharing memory via ``passive'' data objects that can be accessed directly (i.e.~without the overhead of message-passing). We refer to \cite{Nienaltowski07} and \cite{Morandi-NM14a} respectively for more detailed discussions of these concepts.

\begin{figure}[t!]
	\centering
	\includegraphics[width=0.325\textwidth]{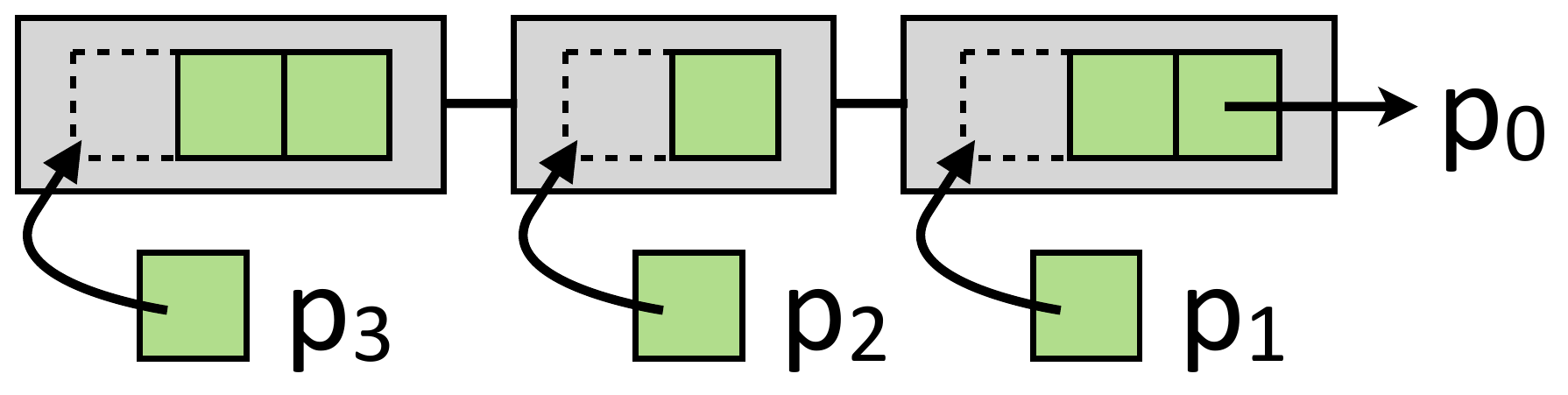}
	\vspace{-2ex}\caption{Three processes $(p_1,p_2,p_3)$ logging requests on another $(p_0)$}\label{fig:qoq_figure}
\end{figure}

\myparagraph{\scoop Runtime.}\label{sec:runtime} The concurrent programming abstractions presented rely on the existence of a runtime that can correctly and efficiently realize them. At the core of \scoop's runtime is a simple execution model for managing requests that are sent between processes. Each process is associated with a ``queue of queues''~\cite{West-NM15b}, that is, a \textsc{fifo} queue itself containing (possibly several) \textsc{fifo} subqueues for storing incoming requests. Each of these subqueues represents a ``private area'' for some other process to log requests, in program text order, and without interference from other processes (since they have their own subqueues). \cref{fig:qoq_figure} visualizes three processes $(p_1,p_2,p_3)$ simultaneously logging requests (green blocks) on another process $(p_0)$. The process $p_0$ is handling the subqueues one-by-one in the order that they were created, and handles the requests within them in the order that they were logged there, hence ensuring the \scoop reasoning guarantees.

Consider again the process that calls \eiffel{transfer (acc1, acc2, 100)} on two separate accounts, \eiffel{acc1} and \eiffel{acc2}. Under the current runtime, the handlers of \eiffel{acc1} and \eiffel{acc2} both generate a private subqueue on which the calling process can log requests (i.e.~the \eiffel{balance} queries and \eiffel{set_balance} commands) without interruption for the duration of the block. Should another process also need to log requests on an account, then a new private subqueue is generated for it and its requests can be logged without waiting.

We remark that earlier versions of the \scoop runtime additionally provided timing guarantees by not allowing processes to enqueue requests concurrently~\cite{Morandi-SNM13a}. A formal comparison with the current semantics is given in~\cite{Corrodi-HP16a}.

\section{Overview of Distributed \scoop}\label{sec:architecture}

In this section we present \dscoop (for Distributed \scoop), which combines network objects and the \scoop synchronization semantics into a single, distributed programming model that maintains the simplicity of the original abstractions. We present an overview of its architecture and communication protocol, and explain how separate calls are generalized to potentially remote objects (Sections~\ref{sec:control} and \ref{sec:compensation} describe in more detail how control of remote objects is achieved in \dscoop, and how the system compensates for unresponsive clients).

A prototype implementation of the \dscoop model is available online~\cite{dscoop:website}. Our prototype builds upon the \scoop support for Eiffel in EiffelStudio~\cite{SCOOP-EiffelStudio-Reference}, which implements the model using threads and shared memory. \dscoop generalizes the implementation, allowing for multiple instances of potentially remote \scoop programs to communicate, under-the-hood, by asynchronous message passing.

\myparagraph{Architecture.} In \dscoop, an instance of a \scoop program is called a \emph{node}. A node can open a connection to another node through a network socket, which is then shared by all of its processes. A node can request the \emph{index object} of another node, which is a user-defined object that typically provides the API of the node, or some form of registry. It is valid for a node to not supply an index object, typically if it is a client in a client-server style setup. To be able to accept incoming connections from other nodes, a node must start a \emph{server} and provide its own index object (or a factory that generates them). Every node in a \dscoop network has a unique \emph{identifier} (ID), which is independent of any other IDs such as IP addresses. Object references in \dscoop include this node ID, along with their object and process identifiers (as in classical \scoop), with the latter important for determining the number of processes involved in a separate block.

The nodes in \dscoop networks communicate, via their connections, using an asynchronous message-passing scheme. Messages conform to a protocol and can be one of two types: a \emph{request}\footnote{Note that these are distinct from the requests used for inter-process communication  in \scoop.} or a \emph{reply}. Requests are sent from a \emph{client} node to a \emph{supplier}, defining work for the supplier to do. Replies are sent back from the supplier to the client indicating the outcome.%

Within nodes, we rely on existing mechanisms of \scoop for garbage collecting local objects and processes. \dscoop however must also account for objects used by multiple nodes. To achieve this, we use a distributed garbage collection algorithm similar to that of Birrell et al.~\cite{Birrell93distributedgarbage}.

\myparagraph{Requests and Replies.} Messages in the \dscoop communication protocol have \emph{subjects} which convey their intended semantics. Messages that are requests can have one of many different subjects which we outline in the following. Replies however only indicate success (\msg{OK}) or failure (\msg{FAIL}), sometimes with additional arguments, such as the result of a query call.

The simplest request subjects are \msg{HELLO}, \msg{PING} and \msg{INDEX}, which respectively initialize a connection between nodes, test whether an existing one is still alive, and request the index object of the supplier node (which typically provides an API of methods for retrieving more objects).

A number of requests are required to realize a separate block involving remote objects. A \msg{PRELOCK} request announces that a process in a client node wishes to log calls on one or more processes in a supplier node. When a supplier is ready, the client can issue a \msg{LOCK} request to announce it is now entering the separate block. Following this, it can issue requests corresponding to asynchronous method calls (\msg{CALL}), synchronous calls (\msg{SCALL}), and queries (\msg{QCALL}). To announce leaving the separate block, the client sends an \msg{UNLOCK} request. (We describe in more detail how these requests establish control in Section~\ref{sec:control}.)

Requests with the subjects \msg{SHARE} and \msg{RELEASE} are respectively used for obtaining and revoking permission for given object references to be shared with third party nodes. They are used by \dscoop for garbage collecting.

Finally, \msg{AWAIT} and \msg{READY} requests are used to implement condition synchronization on remote objects. In short: if the condition does not hold, the client process issues an \msg{AWAIT} request before going to sleep. This instructs the supplier to wake it up with a \msg{READY} request once the state of the remote objects changes, so that the condition can be checked again.

\myparagraph{Message Handling.} Incoming messages are handled by the request handlers of \dscoop nodes in multiple stages, depending on their subjects. If an incoming message has the subject \msg{HELLO}, \msg{PING}, \msg{SHARE}, or \msg{RELEASE}, then it is handled directly. If a message is a reply, then it is relayed to the appropriate process within the node. Messages addressed to other nodes are relayed.

For messages concerning separate blocks and condition synchronization, a more careful treatment is required. In \dscoop, every node has a special designated \emph{proxy process} for handling incoming lock and call requests. Associated with these proxy processes are \emph{proxy objects}, which are surrogates (or placeholders) for actual remote objects, holding references to them. This additional layer is used to catch special contexts in which calls are treated differently. For lack of space we do not go into detail, but mention two of the most important: callbacks (see~\cite{Nienaltowski07}), and a \scoop extension for passive data objects (see~\cite{Morandi-NM14a}).

To minimize the overhead of proxy processes and objects, they are created only when needed and removed when they are not. For example, if not existing already, receiving a \msg{LOCK} request with some given object identifiers will trigger the creation of a proxy process on that node and proxies for those objects. And when no longer in use by local processes, they can be collected by the local \scoop garbage collector.

\myparagraph{Remote Calls in Separate Blocks.} The communication protocol presented is ultimately the glue that allows for network objects to be used within the \scoop framework. Our aim was to make the fusion of these concepts as seamless as possible: programmers should not need to be aware of the communication protocol for network objects, and the core abstractions of \scoop should not need to be fundamentally reinvented to accommodate the extension.

In \dscoop we were able to maintain the original abstractions provided by separate blocks, while also providing a natural generalization to support objects residing on other nodes. When a process needs to make a call on a \eiffel{separate} object, there are now three possible cases to distinguish. If the target object shares the same process (and thus, obviously, the same node), the call is executed immediately---as in \scoop. If the target object has a distinct process but on the same node, the process logs a request in a private subqueue for the caller (see Section~\ref{sec:scoop})---as in \scoop. If the target object has a distinct process on a remote node, however, the \dscoop communication protocol comes into play, and a \msg{CALL} message is sent to to the remote node.

\section{Controlling Remote Objects}\label{sec:control}

We have presented an overview of the \dscoop architecture, its messaging protocol, and its generalization of \eiffel{separate} blocks to support calls on remote objects. In this section, we describe how control of remote objects and thus distributed separate blocks are achieved.

In \dscoop, separate blocks are handled in three phases: (i) the \emph{prelock phase}, for ensuring a correct ordering; (ii) the \emph{issuing phase}, for enqueuing calls; and (iii) the \emph{execution phase}, for executing calls. The issuing phase happens strictly after the prelock phase. While the execution phase cannot start before the issuing phase, the two can otherwise overlap due to asynchronicity.

\myparagraph{Prelock Phase.} In standard \scoop, if a process enters a separate block, the processes handling the \eiffel{separate} objects generate private subqueues for logging calls (see Section~\ref{sec:scoop} and Figure~\ref{fig:qoq_figure}). In \dscoop however, if a process enters a separate block involving \eiffel{separate} objects on remote nodes, messages must be sent to trigger the generation of subqueues in a way that preserves the usual reasoning guarantees. We refer to this messaging phase as the \emph{prelock phase}.

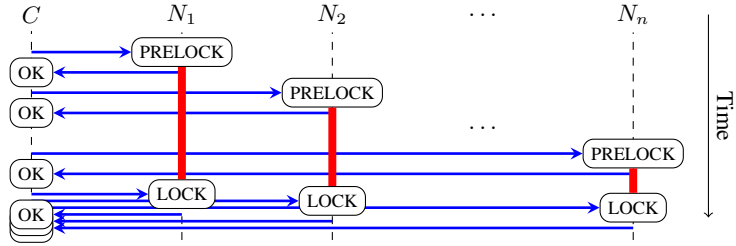
\begin{figure}[t!]
\centering
\begin{tikzpicture}[
message/.style={
	rounded corners, 
	fill=white
},
com/.style={
	line width=1pt, 
	blue, 
	>=stealth
},
]

\newcounter{msgno}
\newlength{\msgpos}
\pgfmathsetlength{\msgpos}{-.5cm}

\newcommand{\scoopmessage}[6][\themsgno] {
	\pgfmathaddtolength{\msgpos}{#2}
	\node[draw, message] (#1) at (#4,\msgpos) {\scriptsize#5};
	\draw[->,com] (#3,\msgpos) -- (#1);
	\stepcounter{msgno};
}

	\newlength{\msggap}
	\setlength{\msggap}{-0.09cm}

\node[fill=white] (n0) at (0,0) {$C$};
\node[fill=white] (n1) at (2,0) {$N_1$};
\node[fill=white] (n2) at (4,0) {$N_2$};
\node[fill=white] (nn) at (8,0) {$N_n$};
\path[draw, dashed] (0,-3) -- (n0);
\path[draw, dashed] (2,-3) -- (n1);
\path[draw, dashed] (4,-3) -- (n2);
\path[draw, dashed] (8,-3) -- (nn);
\draw[->] (9,0) -- (9,-2.7) node[pos=.5, sloped, above] {Time};

\scoopmessage[pre1]{0\msggap}{0}{2}{PRELOCK}

\scoopmessage{3\msggap}{2}{0}{OK}

\scoopmessage[pre2]{3\msggap}{0}{4}{PRELOCK}

\scoopmessage{3\msggap}{4}{0}{OK}

\scoopmessage[pre3]{6\msggap}{0}{8}{PRELOCK}

\scoopmessage{3\msggap}{8}{0}{OK}

\scoopmessage[lock3]{5\msggap}{0}{8}{LOCK}

\scoopmessage[lock2]{-1\msggap}{0}{4}{LOCK}

\scoopmessage[lock1]{-1\msggap}{0}{2}{LOCK}

\scoopmessage{5\msggap}{8}{0}{OK}

\scoopmessage{-1\msggap}{4}{0}{OK}

\scoopmessage{-1\msggap}{2}{0}{OK}

\node[fill=white] at (6,0) {$\ldots$};
\node[fill=white] at (6,-1.5) {$\ldots$};

\path[draw, line width=3pt, color=red] (pre1) -- (lock1);
\path[draw, line width=3pt, color=red] (pre2) -- (lock2);
\path[draw, line width=3pt, color=red] (pre3) -- (lock3);
\end{tikzpicture}
	\vspace{-2ex}
\caption{Prelock phase: a process on node $C$ is entering a separate block involving \eiffel{separate} objects on remote nodes $N_1,\dots N_n$}
\label{fig:locking}
\end{figure}

A client node seeking to enter a separate block involving remote objects must first announce its intention by sending \msg{PRELOCK} requests to the nodes they reside on. This is done in a fixed order (a global order based on node IDs) to avoid deadlocks, and one-at-a-time; an \msg{OK} reply must be received before the next \msg{PRELOCK} is sent. Once the last such request is successful, the client node announces that it is entering the separate block and will start issuing calls. This announcement is made via \msg{LOCK} requests, which can be sent asynchronously in any order. By replying with \msg{OK}, the supplier nodes are acknowledging that the involved processes have created private subqueues and are ready to enqueue calls from the client. Figure~\ref{fig:locking} exemplifies this phase for a client node $C$ that wishes to enter a separate block involving remote objects on supplier nodes $N_1,\dots N_n$. Here, an arrow denotes the transmission of a message, with its subject given at the end (additional parameters are not visualized).

When multiple nodes are entering prelock phases involving common supplier nodes, blocking must occur in order to maintain the separate block order guarantees. In particular, if a \msg{PRELOCK} message is sent but the supplier is already involved in the prelock phase of a competing node, then the system blocks on that message. Instead of blocking for the whole of the competing node's separate block, \dscoop permits a more fine-grained and efficient solution. In particular, it only blocks until the competing node leaves its prelock phase and starts issuing calls. That is to say, \dscoop only blocks while ``setting up'' the subqueues in a correct order; competing issuing phases can otherwise safely run concurrently.

\myparagraph{Issuing and Execution Phases.}
The prelock phase ends and the issuing phase begins when the final \msg{LOCK} request is successful. At this point, the processes handling all the involved remote objects are ready to enqueue calls. In most circumstances, commands on remote objects are requested via asynchronous \msg{CALL} messages, and queries are requested via synchronous \msg{QCALL} messages. The supplier nodes enqueue commands and immediately reply with an \msg{OK}. When a query is received however, the supplier node enqueues it, but only replies once it has been executed (passing the result in an additional parameter of the \msg{OK} message).

The execution phase begins with the execution of the first logged call. If all the calls are asynchronous, it can take place strictly after the issuing phase. The issuing phase ends on sending the \msg{UNLOCK} message; the execution phase ends on processing it.

\begin{listing}[t!]
	\centering
  	\begin{eiffelcode}
  		withdraw (s: separate ACCOUNT; am: NATURAL)
  		do
	  		if s.balance >= am then
		  		s.set_balance (s.balance - am)
	  		else
				-- Notify user
	  		end
  		end
  	\end{eiffelcode}
  	\vspace{-3ex}
	\caption{Bank account withdrawal method in \dscoop}
	\label{lst:withdraw}
\end{listing}

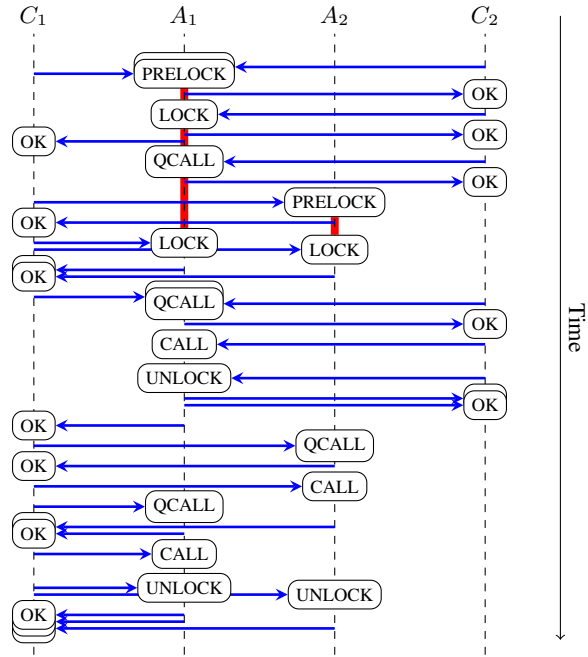
\begin{figure}[t!]
	\centering
	\begin{tikzpicture}[
	message/.style={
		rounded corners, 
		fill=white
	},
	com/.style={
		line width=1pt, 
		blue, 
		>=stealth
	},
	]
	
	\setcounter{msgno}{0}
	\pgfmathsetlength{\msgpos}{-.5cm}
	
	\newcommand{\scoopmessage}[6][\themsgno] {
		\pgfmathaddtolength{\msgpos}{#2}
		\node[draw, message] (#1) at (#4,\msgpos) {\scriptsize#5};
		\draw[->,com] (#3,\msgpos) -- (#1);
		\stepcounter{msgno};
	}

	\setlength{\msggap}{-0.09cm}
	
	\node[fill=white] (c1) at (0,0) {$C_1$};
	\node[fill=white] (a1) at (2,0) {$A_1$};
	\node[fill=white] (a2) at (4,0) {$A_2$};
	\node[fill=white] (c2) at (6,0) {$C_2$};
	
	\path[draw, dashed] (0,-8.5) -- (c1);
	\path[draw, dashed] (2,-8.5) -- (a1);
	\path[draw, dashed] (4,-8.5) -- (a2);
	\path[draw, dashed] (6,-8.5) -- (c2);
	\draw[->] (7,0) -- (7,-8.3) node[pos=.5, sloped, above] {Time};
		
	\scoopmessage[pre1]{2\msggap}{6}{2}{PRELOCK}
	
	\scoopmessage[pre2]{\msggap}{0}{2}{PRELOCK}
	
	\scoopmessage{3\msggap}{2}{6}{OK}
	
	\scoopmessage[l1]{3\msggap}{6}{2}{LOCK}	
	
	
	\scoopmessage{3\msggap}{2}{6}{OK}
	
	\scoopmessage{\msggap}{2}{0}{OK}
	
	\scoopmessage{3\msggap}{6}{2}{QCALL}
	
	\scoopmessage{3\msggap}{2}{6}{OK}
	
	\scoopmessage[pre4]{3\msggap}{0}{4}{PRELOCK}
	
	\scoopmessage{3\msggap}{4}{0}{OK}
	
	\scoopmessage[l3]{4\msggap}{0}{4}{LOCK}
	
	\scoopmessage[l4]{-1\msggap}{0}{2}{LOCK}
	
	\scoopmessage{4\msggap}{2}{0}{OK}
	
	\scoopmessage{1\msggap}{4}{0}{OK}
	
	\scoopmessage{3\msggap}{0}{2}{QCALL}
	
	\scoopmessage{1\msggap}{6}{2}{QCALL}
	
	\scoopmessage{3\msggap}{2}{6}{OK}
	
	\scoopmessage{3\msggap}{6}{2}{CALL}
	
	\scoopmessage{5\msggap}{6}{2}{UNLOCK}
	
	\scoopmessage{3\msggap}{2}{6}{OK}
	
	\scoopmessage{1\msggap}{2}{6}{OK}
	
	\scoopmessage{3\msggap}{2}{0}{OK}
	
	\scoopmessage{3\msggap}{0}{4}{QCALL}
	
	\scoopmessage{3\msggap}{4}{0}{OK}
	
	\scoopmessage{3\msggap}{0}{4}{CALL}
	
	\scoopmessage{3\msggap}{0}{2}{QCALL}
	
	\scoopmessage{3\msggap}{4}{0}{OK}
	
	\scoopmessage{1\msggap}{2}{0}{OK}
	
	\scoopmessage{3\msggap}{0}{2}{CALL}
	
	\scoopmessage{6\msggap}{0}{4}{UNLOCK}
	
	\scoopmessage{-1\msggap}{0}{2}{UNLOCK}
	
	\scoopmessage{6\msggap}{4}{0}{OK}
	
	\scoopmessage{-1\msggap}{2}{0}{OK}
	
	\scoopmessage{-1\msggap}{2}{0}{OK}
	
	  \begin{pgfonlayer}{background}
	\path[draw, line width=3pt, color=red] (pre1) -- (l1);
	\path[draw, line width=3pt, color=red] (l1) -- (l4);
	\path[draw, line width=3pt, color=red] (pre4) -- (l3);
	  \end{pgfonlayer}
	\end{tikzpicture}
	\vspace{-2ex}
	\caption{All three phases: a process on $C_1$ calls \eiffel{transfer} on $A_1$ and $A_2$; a process on $C_2$ concurrently calls \eiffel{withdraw} on $A_1$}
	\label{fig:communicationexample}
\end{figure}

\myparagraph{Example Communication.}
We return to our running bank account example, which we extend with a simple method \eiffel{withdraw} (Listing~\ref{lst:withdraw}) for withdrawing a given amount from a given account that we assume to be remote. The method first synchronously queries the remote object to check that the balance is sufficient, before asynchronously decreasing the balance.

Suppose we have a running \dscoop system with two bank accounts on different nodes $(A_1,A_2)$. Suppose now that a client node $(C_1)$ is trying to \eiffel{transfer} an amount from $A_1$ to $A_2$, while another client node $(C_2)$ is trying to \eiffel{withdraw} an amount from $A_1$. Recall that the bodies of both methods are separate blocks (involving, respectively, \eiffel{separate} accounts on $A_1,A_2$ and $A_1$). \Cref{fig:communicationexample} visualizes the messages exchanged in one possible behavior.

Observe that both clients initially send a \msg{PRELOCK} request to $A_1$. The request from $C_2$ is received first and is therefore answered immediately; meanwhile, $C_1$ blocks. Since $C_2$ only seeks control over a process on $A_1$, it proceeds to send a \msg{LOCK} request, thus completing its prelock phase and generating its private subqueue on $A_1$. This allows $C_1$ to unblock and its first \msg{PRELOCK} request finally succeeds.

Since the prelock phase of one client can take place in parallel to the issuing and execution phases of another, $C_2$ already starts issuing calls before $C_1$ concludes its prelock phase. In particular, it requests the \eiffel{balance} query (via \msg{QCALL}) which is executed synchronously (and the balance amount returned). Following this, $C_1$ requests a \msg{PRELOCK} on $A_2$ (which is uncontended), before completing its prelock phase by sending \msg{LOCK} requests to $A_1$ and $A_2$.

At this point, both $C_1$ and $C_2$ issue \eiffel{balance} queries (\msg{QCALL})---the former is evaluating its conditional guard, and the latter is evaluating the expression in the input of \eiffel{s.set_balance (s.balance - am)}. Since $C_2$ completed its prelock first, its private subqueue on $A_1$ is ahead of the subqueue for $C_1$, and so its call is executed first. Following this, $C_2$ requests an asynchronous command (\msg{CALL}) to update the balance, and then exits its separate block via an \msg{UNLOCK} request. Once acknowledged, $C_2$ knows that the whole transaction (\eiffel{balance} and then \eiffel{set_balance}) was successful, and its effects become visible to other clients. Once the \msg{OK} corresponding to its earlier \msg{QCALL} arrives, $C_1$ can resume issuing the remaining calls in its separate block before exiting via \msg{UNLOCK} requests to $A_1$ and $A_2$.

Note that the reasoning guarantees of the separate blocks have been maintained. The calls are executed in program text order and without intervening calls from other nodes: within a separate block, multiple \eiffel{balance} calls in sequence thus always return the same result. The combination of the prelock phase and the underlying queue of queues semantics prevents the possibility of interleavings that break this.

\section{Compensating for Failure}\label{sec:compensation}

Our presentation of \dscoop has thus far focused on the challenge and intricacies of combining the network objects abstraction with a concurrency model and runtime. In this section, we turn our attention to a topic that cannot be ignored in the setting of distributed computing: coping with failure.

While failure can often be managed simply---a fixed timeout is used, for example, to manage it in prelock phases---failure in the middle of a separate block, when only some of the side-effecting commands have been issued, needs a more elaborate solution. We introduce \emph{compensation}, \dscoop's mechanism for reacting to such failure, and demonstrate its use on our running example.

\myparagraph{Compensation.} In \dscoop, upon failure of a supplier, the client is informed using exceptions, and can react to it appropriately in a \eiffel{rescue}-clause. However, the suppliers in separate blocks are in general oblivious to the status of the client. Our solution is to introduce \emph{compensation}, a supplier-side mechanism for reacting to client nodes that become unresponsive or disconnect prematurely. The technique registers user-provided closures on suppliers that, before releasing objects controlled by disconnected clients, are executed to restore consistency.

The basic technique is adapted from well-established usage in transactions, in particular, for recovering from long-running transactions or transactions with side effects. It fits naturally with the \dscoop model, given that separate blocks are transaction-like in the sense that other clients cannot observe the \eiffel{separate} objects in intermediate states. One can think of a \msg{LOCK} and \msg{UNLOCK} pair as being the beginning and end of a transaction; after \msg{UNLOCK} is acknowledged, all changes become visible.

The scope of compensation is the issuing phase, and encompasses all executed calls on processes that have been acquired during the prelock phase (and only those processes). In the case of nested separate blocks, the outer block has to take into account that the effects of the inner block are already visible if an \msg{UNLOCK} was issued. This is different to most definitions of nested transactions, in which the inner transaction always finishes together with the outer transaction.

\myparagraph{Defining Compensation.}
Compensation closures are provided by the user as the input of special methods for registering compensation. (We remark that closures are given with the Eiffel keyword \eiffel{agent}, and can refer to existing methods.) It is possible to define them in the client or the supplier. A client-defined compensation closure is registered before the call to the method to be compensated (and is ignored by the supplier if no request follows). A supplier-defined compensation closure is provided within the called method. The latter comes with the advantage that compensation is defined together with the method, but the former allows for more flexibility: different compensations can be defined depending on where the call is made, which is particularly useful for methods that do not always need compensation.

\begin{listing}[t!]
	\caption{A \eiffel{set_balance} method together with possible compensation}
	\label{lst:compensation}
	\begin{subfigure}[t]{.5\textwidth}
		\caption{Client-defined compensation}
		\label{lst:compensation:client}
		\begin{eiffelcode}
...
	t.compensate (agent 
		t.set_balance (t.balance))
	t.set_balance (t.balance + am)
...
		\end{eiffelcode}
	\end{subfigure}		
	\begin{subfigure}[t]{.5\textwidth}
		\caption{Supplier-defined compensation}
		\label{lst:compensation:supplier}
		\begin{eiffelcode}
set_balance (nb: NATURAL)
	do
		compensate (agent 
			set_balance (balance))
		balance := nb
	end
		\end{eiffelcode}
	\end{subfigure}		
\end{listing}

Consider the simple method \eiffel{set_balance} for bank account objects (\Cref{lst:compensation}) which sets the \eiffel{balance} of an account to some provided input. The listing also includes examples of how to make it compensable. On the left is a snippet of the body of \eiffel{transfer}, now annotated with client-defined compensation before the call. On the right is supplier-defined compensation, provided at the beginning of the method body. In both cases, the \eiffel{balance} argument to the closure (\eiffel{agent}) is evaluated to the original balance, so it will restore the old balance if called.

\myparagraph{Implementing Compensation.}
Upon receiving a \msg{LOCK} request, a supplier node stores the IDs of the newly requested processes in a stack. This stack is mainly used to identify which processes need to be released upon \msg{UNLOCK}. Each of the process entries also contains a reference to a set of compensation closures, extracted from the program text. These closures are accompanied by relative timestamps, so that within all the sets for this client each number is unique and a later registration has a strictly higher number than an earlier one. Whenever a process is unlocked normally (i.e.~not due to premature disconnection) the respective set is cleared. However, if a client node disconnects prematurely, all sets associated with the client are merged and then ordered by the timestamps. The execution of the compensation closures is done in reverse order.

\begin{figure}[t!]
\centering
\begin{tikzpicture}[
y=-1cm,
x=1cm,
element/.style={
		draw,
	    node distance=0,
	    outer sep=0pt,
   		align=center,
        fill=white
    },
call/.style={
		font=\scriptsize,
		fill=green!90,
	    node distance=1mm,
	    outer sep=0pt,
   		minimum height=12
    },
call2/.style={
   		font=\scriptsize,
   		fill=blue!30,
	    node distance=1mm,
	    outer sep=0pt,
      		minimum height=12
    },
com/.style={
        line width=1pt, 
        blue, 
        >=stealth
    },
    nodelist direction/.is choice,
    nodelist direction/.default=horizontal,
    nodelist direction/horizontal/.style={
        start chain=going right,
    },
    nodelist direction/vertical/.style={
        start chain=going below, 
    },
    nodelist/.style={
	    node distance=0,
	    outer sep=0pt,
        nodelist direction=#1,
        every node/.append style={
            on chain,
            draw
        },
    }
]
\node[call,minimum width=150] (a) at (0,0) {\eiffel{a}};
\node[call,minimum width=100,above=of a,] (c) {\eiffel{c}};
\node[call,minimum width=20,left=of c,] (b) {\eiffel{b}};
\node[call,minimum width=20,right=of c,] (d) {\eiffel{d}};
\node[call2,minimum width=50,above=of c,yshift=3mm] (f) {\eiffel{f}};
\node[call2,minimum width=20,left=of f] (e) {\eiffel{e}};
\node[call2,minimum width=20,right=of f,] (g) {\eiffel{g}};
\node[call,minimum width=40,right=of a,] (h) {\eiffel{h}};

\node[node distance=2mm, outer sep=0, left=of a, rotate=90, xshift = 15] (label1) {P 1};
\node[draw,fit=(label1) (a) (b) (d) (h)] (box1) {};

\node[node distance=2mm, outer sep=0, left=of e, rotate=90, xshift = 9] (label2) {P 2};
\node[draw,fit=(label2) (e) (g)] (box2) {};

\node[fit=(box1) (box2)] (box) {};

\draw[decorate,decoration={brace,amplitude=10pt}] (box.south west) -- (box.north west) node [left, rotate=90, xshift=-10, yshift=15] {Call stack};

\end{tikzpicture}
\vspace{-2ex}
\caption{Example call stack}
\label{fig:transactions}
\end{figure}
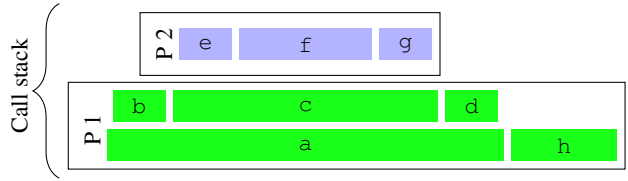

\Cref{fig:transactions} shows the call stack caused by a remote client calling the method \eiffel{a} and then \eiffel{h}. The targets of \eiffel{a}, \eiffel{b}, \eiffel{c}, \eiffel{d} and \eiffel{h} are owned by process $P_1$, while the targets of the calls \eiffel{e}, \eiffel{f}, and \eiffel{g} are owned by process $P_2$. During the execution of \eiffel{c}, $P_1$ acquires control over $P_2$ to execute. After \eiffel{a} is finished, the client sends another request to execute \eiffel{h} before releasing $P_1$.

We now take a look at three failure scenarios, all of them due to a premature disconnect by the client. If the client disconnects before \eiffel{a} is executed, nothing happens. The client's control over $P_1$ is simply lifted. The second case is more complex: if the client disconnects while \eiffel{a} is executing, the calls \eiffel{a}, \eiffel{b}, \dots \eiffel{g} are all executed as requested. Since $P_1$ is issuing the \msg{UNLOCK} request to $P_2$ before finishing itself, the changes done by \eiffel{e}, \eiffel{f}, \eiffel{g} are visible. The disconnect then causes the compensation closures of \eiffel{d}, \eiffel{c}, \eiffel{b}, \eiffel{a} to be executed before control over $P_1$ is released. Consequently, the compensation of \eiffel{c} has to deal with the fact that the changes due to \eiffel{e}, \eiffel{f}, \eiffel{g} are already visible.

If the client issued the call to \eiffel{h} but got lost before sending the \msg{UNLOCK} request, the situation is similar, with the one difference being that the compensation of \eiffel{h} is executed before the others.

\section{Evaluation}\label{sec:evaluation}
We evaluated \dscoop against Java RMI to gauge its performance against a well-established and widely used approach based on network objects. We sought to collect evidence towards answering two questions. First, is there a performance overhead associated with the automatic synchronization in \dscoop, and does it become incommensurate with the effort to manually write synchronization code? Second, do the language abstractions of \dscoop facilitate simpler code?

\myparagraph{Example Selection.}
\dscoop and Java RMI have many differences: not only in the model, but also in terms of the underlying programming languages (Eiffel and Java) which have many points of variation regarding performance and compilers. In this context, we devised a set of four microbenchmarks isolated to comparing the performance of calls: (i) \emph{command call}, in which a single client sends a series of command calls to the supplier; (ii) \emph{query call}, analogous, but with query calls; (iii) \emph{locking and command call}, in which a few clients compete to control a supplier object and send a single command call; and (iv) \emph{locking and query call}, analogous, but with a single query call.

In addition to microbenchmarks, we also evaluated \dscoop against Java RMI on three larger examples. First, \emph{dining philosophers}, a classical example where multiple objects (forks) are repeatedly controlled. For this benchmark, all philosophers and forks reside on different nodes, and we assume that eating, using the fork, and thinking take no time. Second, a more practical example: a \emph{log server}, in which various events are logged. Here, there are multiple log servers for redundancy, meaning that copies of logs can still be retrieved if one fails. To ensure a consistent ordering across servers, a client must control all of them before adding the entries. In our benchmark, three clients repeatedly generate a simple log message, gain control across the servers, and then place it. Third, a \emph{pipeline} representing distributed services. Each stage waits until the previous stages are ready before retrieving data and processing it. Each stage provides one operation of the well known formula $\sqrt{a^2+b^2}$. We measured the time the final stage needed for a specific number of calculations.

For Java RMI, explicit locking was used to establish a comparable flexibility in the clients. Furthermore, the Java code explicitly orders the locks so as to avoid deadlocks. The source code of the examples and of \dscoop itself can be found on our supplementary material webpage~\cite{dscoop:website}.

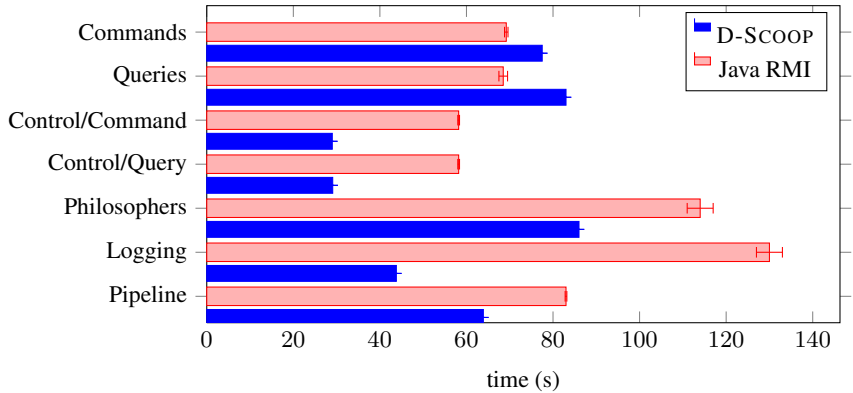
\begin{figure}[t!]
\footnotesize
\begin{tikzpicture}
\begin{axis}[
	xbar,
	xmin=0,
	xlabel={time (s)},
	symbolic y coords={
		Pipeline,
		Logging,
		Philosophers,
		Control/Query,
		Control/Command,
		Queries,
		Commands,
		},
	ytick=data,
	width=10cm,
	height=5.8cm,
	ylabel=]
	\addplot+[
		bar width=6pt,
		fill=blue,
		error bars/.cd, 
		x dir=both,x explicit,] coordinates
	{
		(77.5,Commands) +- (0,0)
		(83.0,Queries) +- (0,0)
		(29.0,Control/Command) +- (0,0)
		(29.1,Control/Query) +- (0,0)
		(86.0,Philosophers) +- (0,0) 
		(43.8,Logging) +- (0,0) 
		(63.9,Pipeline) +- (0,1)
	}; 
	\addplot+[
		bar width=7pt,
		error bars/.cd, 
		x dir=both,x explicit,] coordinates
	{
		(69.2,Commands) +- (0.4,0)
		(68.5,Queries) +- (1.0,0)
		(58.2,Control/Command) +- (0.2,0)
		(58.2,Control/Query) +- (0.2,0)
		(114,Philosophers) +- (3,0)
		(130,Logging) +- (3,0)
		(83.0,Pipeline) +- (0.2,0)
	}; 
	\addplot+[
	bar width=7pt,
	error bars/.cd, 
	x dir=both,x explicit,] coordinates
	{
	}; 
	\legend{\dscoop, Java RMI, Java RMI opt}
\end{axis}
\end{tikzpicture}
\vspace{-2ex}
\caption{Benchmark results: each run involved several thousand iterations (see~\cite{dscoop:website})}
\label{fig:benchmarks}
\end{figure}

\myparagraph{Performance.}
Overall, we found that despite the potential overhead of automatic synchronization, \dscoop's performance is competitive with---and can be superior to---explicit locking-based synchronization in Java RMI. The results of the performance evaluation are listed in \cref{fig:benchmarks} and are the averages of 30 runs; we used two off-the-shelf laptops connected by an ethernet cable. The microbenchmarks show that the performance of both \dscoop and Java RMI is similar when just issuing commands or queries. \dscoop commands are a bit quicker than \dscoop queries due to them being asynchronous, whereas in RMI both are synchronous. When it comes to the control microbenchmarks, the built-in synchronization in \dscoop allows for a more significant improvement in speed, both for synchronous and asynchronous calls. However, the synchronization overhead prevents the asynchronous advantage of Control/Command translating into faster performance than Control/Query.

For both the dining philosophers and the logging example, the fact that the prelock phase can be done in parallel with the issuing and execution phase of another client proves to be a significant advantage in comparison to RMI. In addition, the logging example shows the advantage of asynchronous calls in \dscoop. The underlying semantics make it possible to ensure control over multiple nodes and have multiple clients issuing asynchronous calls at the same time. The pipeline example has less congestion around the protected objects; here, the advantage of \dscoop lies solely in slightly fewer messages sent due to more powerful synchronization mechanics. 

\myparagraph{Simplicity.} 
Our second question asked whether the language abstractions also yield simpler code. For our seven benchmarks, we recorded: (i) the number of classes involved, excluding primitive types, classes, and strings, and ignoring the RMI remote interface; (ii) the number of features (i.e.~attributes and methods), ignoring the Java ``getters'' in RMI since they just return an otherwise counted attribute; and (iii) the number of written instructions, excluding boilerplate code. This ensures that the differences are only due to synchronization. Table~\ref{tab:instructions} lists the results.

\begin{table}[t!]
	\caption{Code complexity}
	\label{tab:instructions}
	\centering
	\begin{tabular}{l|cccccc}
		& \multicolumn{2}{c}{Classes} & \multicolumn{2}{c}{Features} & \multicolumn{2}{c}{Instructions} \\ 
		& RMI & \dscoop & RMI & \dscoop & RMI & \dscoop \\ 
		\hline
		Microbenchmarks & 3 & 2 & 8 & 6 & 19 & 13 \\ 
		Dining philosophers & 3 & 2 & 6 & 3 & 18 & 10 \\ 
		Logging & 6 & 3 & 16 & 9 & 23 & 10 \\ 
		Pipelines & 2 & 1 & 10 & 16 & 62 & 42 \\ 
	\end{tabular} 
\end{table}

As can be seen, the solutions in \dscoop are much more compact across the three measurements. In the case of advanced techniques such as condition synchronization---an in-depth discussion is omitted for brevity---the complexity of RMI increases further still. Note that not included in the RMI examples are compensation and the automatic releasing of locks, since they are difficult to achieve in that framework. Also, although the usage of a lock or semaphore is counted as a class, its features are not counted in the feature column since they are already provided by the library. We remark that these numbers only indicate that \dscoop programs are more compact than their RMI counterparts. What we leave to future work is a study of users themselves to determine whether the \dscoop abstractions are easier to read and program with, regardless of their compactness. (An existing \scoop study is encouraging~\cite{Nanz-et_al11a}.)

\section{Related Work}\label{sec:related}

There is a wide selection of work addressing concurrency and distribution in the object-oriented paradigm. Here, we highlight some work that is closest to our own.

The active object~\cite{Lavender-Schmidt96a} design pattern (which inherits from the actor model~\cite{Agha86a}), like \scoop, decouples method calls from method executions. Such objects are associated with their own processes, which can send messages to each other asynchronously, introducing concurrency. Despite the similarity to \scoop, active objects lack the guarantee of interference-freedom when multiple objects are involved. Furthermore, non-active objects have to be protected manually, and there is no built-in support for condition synchronization (although it is possible to use the observer pattern to actively notify waiting processes). \scoop can be seen as an advanced form of active objects: objects are by default active, but multiple objects can share the same process. In addition, the \scoop synchronization mechanisms ensure the absence of intervening calls and also protect non-active objects~\cite{Morandi-NM14a}. Condition synchronization is simple (via method preconditions) and does not require signaling.

There have been some successful attempts to generalize ideas from active objects and the actor model to distributed programming frameworks, with some prominent examples including Creol~\cite{Johnsen-Owe-Yu06a}, AmbientTalk~\cite{Dedecker-et_al06a}, and JCoBox~\cite{Schaefer-PH10a}. The latter partitions the object space into ``coboxes'', each with a common thread of control to improve safety; an approach similar to the processes of \scoop and \dscoop. Caromel et al.~\cite{Caromel09asynchronoussequential} consider a way of unifying threads and objects to support simpler reasoning about distributed computing, and provide a formal calculus. An important distinction of \dscoop in comparison to other frameworks is the impossibility of interrupting requests sent to multiple (potentially distributed) objects controlled by different threads, giving the model its transaction-like semantics.

Network objects~\cite{Birrell:1993} share some similarities with active objects, although calls to them are traditionally synchronous to mimic standard method calls, and calls to local network objects are usually handled by the calling process. Creol exemplifies different synchronization approaches possible with active objects, and their natural extension to network objects. Some languages, such as E~\cite{Miller05concurrencyamong}, avoid blocking entirely to ensure deadlock-freedom. This, in our view, can lead to complex behavior that is difficult to understand from the point of view of classical sequential programming. By making synchronization simpler to use, \dscoop potentially reduces (but does not eliminate) the risk of deadlocks.

For dealing with failures, the programming language Argus~\cite{Liskov:1988:Argus} supports ``atomic objects'' that can be used in a transaction. In contrast, our compensation approach is not limited to pure data-objects.

\section{Conclusion} \label{sec:conclusion}
This paper made a case for combining network objects with synchronization models. We presented \dscoop, a distributed programming model obtained by combining the network objects abstraction with the runtime semantics of the object-oriented concurrency model \scoop. We presented an efficient two-phase locking protocol that generalized the strong reasoning guarantees of \scoop to network objects, allowing for interference-free and transaction-like reasoning on (potentially multiple) remotely located objects, without the programmer having to explicitly manage their synchronization. Furthermore, we proposed a compensation mechanism by which \dscoop programs can recover from failure. The evaluation of our prototype implementation \cite{dscoop:website} suggested that \dscoop remains competitive against---and can outperform---explicit locking-based synchronization in Java RMI, a well-established realization of network objects, with the automatic synchronization mechanisms also allowing for more compact code.

In future work, we plan to improve the efficiency of \dscoop with respect to intra-object parallelism~\cite{Johnsen-et_al09a,Henrio-et_al13a}. We will investigate concepts such as slicing~\cite{Schill-Nanz-Meyer13a}, and the possible integration of software transactional memory~\cite{Eugster-Vaucouleur06a}. We will also investigate whether performance can be improved, by (safely) relaxing the requirement that one node communicates with another via a single connection. Finally, we want to formalize the \dscoop semantics using~\cite{Corrodi-HP16a} to test extensions, and provide a formal proof that the protocol and algorithms correctly generalize the \scoop guarantees. \\

\noindent\emph{Acknowledgements.} We thank Sebastian Nanz for his invaluable support throughout this project. We also thank Carlo A.~Furia and the anonymous referees for their helpful comments and criticisms. The underlying research was partially funded by ERC Grant CME \#291389.

\bibliographystyle{splncs03}
\bibliography{references}{}

\end{document}